\documentclass[12pt]{article}
 \usepackage[cp1251]{inputenc} %- для WiN
 \usepackage[english, russian]{babel} %- для рус-англ. переноса
 \usepackage{amsmath, latexsym, amssymb, bm, array, graphics, eucal,
 amsfonts,}%mathscr

 
\usepackage[dvips]{graphicx}

\begin{document}
\thispagestyle{empty}
\large
\renewcommand{\abstractname}{Abstract}
\renewcommand{\refname}{\begin{center}
 REFERENCES\end{center}}
\newcommand{\mc}[1]{\mathcal{#1}}
\newcommand{\E}{\mc{E}}
\makeatother
\begin{center}
\bf  Boundary problems for one-dimensional kinetic equation with
constant collision frequency
\end{center} \medskip

\begin{center}
  \bf
  A. L. Bugrimov\footnote{$fakul-fm@mgou.ru$},
  A. V. Latyshev\footnote{$avlatyshev@mail.ru$} and
  A. A. Yushkanov\footnote{$yushkanov@inbox.ru$}
\end{center}\medskip

\begin{center}
{\it Faculty of Physics and Mathematics,\\ Moscow State Regional
University, 105005,\\ Moscow, Radio str., 10A}
\end{center}\medskip

\begin{abstract}
For the one-dimensional linear kinetic
equation  analytical solutions of prob\-lems about
temperature jump and weak evaporation (condensation) over
flat surface are received. The equation has integral of collisions
BGK (Bhatnagar, Gross and Krook) and constant frequency of
collisions of molecules.
Distribution of concentration, mass speed and temperature is
received.


\medskip


\end{abstract}

\begin{center}
\bf  Introduction
\end{center}

In work \cite {1} the linear one-dimensional kinetic
equation with integral of collisions BGK (Bhatnagar, Gross and
Krook) and frequency of colli\-si\-ons, affine depending on the module
velocity of molecules has been entered.
Preservation laws of
numerical density (concentration) of mole\-cu\-les, momentum of molecules and
energy have been thus used.

In \cite {1} the theorem about structure of general solution
of the entered equation has been proved.

In work \cite {2}, being continuation of \cite{1}, are received
exact solutions of the problem on temperature jump and weak evaporation
(condensation) in rarefied gas for kinetic equation with
 frequency of colli\-si\-ons, affine depending on the module
velocity of molecules.

In the present work which is continuation of \cite{1} and \cite{2},
exact solutions of the problem about temperature jump and weak evaporation
(conden\-sa\-tion) in the rarefied gas are received.
Here the one-dimensional equation with constant frequency
of collisions is used. This equation is a special case of
the kinetic equation with frequency of
collisions, affine depending on the module velocity of molecules.

These two problems following
\cite{3} we will name the generalized Smo\-lu\-chowsky' problem , or simply
the Smoluchow\-sky problem.

Let us stop on history of exclusively analytical solutions
of the genera\-li\-zed Smoluchowsky' problem.

For simple (one-nuclear) rarefied gas with a constant
frequency of collisions of molecules the analytical solution
 of the generalized of Smo\-lu\-chow\-sky' problems it is received in \cite{4}.

In \cite {5} the generalized of Smolu\-chow\-sky' problem  was analytically
solved for simple rarefied gas with frequency of collisions
the molecules,  linearly depending on the module of molecular velocity.
In \cite {6} the problem about strong evaporation (condensation)
with constant frequency of collisions has been analytically solved.

Let us notice, that for the first time the problem
about temperature jump with
frequency of collisions of molecules, linearly
depending on the module
molecular velocity, was analytically solved
by Cassel and Williams in work \cite{7} in 1972.

Then in works \cite{8,9,10} the generalized Smoluchowsky' problem
also analytical solution for case of multinuclear (molecular)
gases has been received.

In works \cite{11,12,13} the problem  about behaviour of the quantum
Boze-gas at low temperatures (similar to the temperature jump problem
for electrons in metal) is considered. We used
the kinetic equation with excitation fonons agrees to N.N. Bogolyubov.

In works \cite{14,15} the problem about temperature jump
for electrons of degenerate plasmas  in metal has been solved.

In work \cite{16} the analytical solution of the  Smoluchowsky' problem
for quantum gases it has been received.

In work of Cercignani and Frezzotti \cite{17} the Smoluchowsky'  problem
it was considered with use of the one-dimensional
kinetic equations. The full analytical solution of
Smoluchowsky' problem with use of Cercignani---Frezotti equation it has
been received in work \cite{18}.

At the same time there is an unresolved problem about temperature jump
and concentration with use of the BGK--equation with
arbitrary dependence of frequency on velocity, in spite of
on obvious importance of the decision of a problem in similar statement.

In the present work attempt to promote in this direction
is made. Here the case of the affine
dependence of collision frequency on molecular velocity in
models of one-dimensional gas is considered.
Model of one-dimensional gas
gave the good consent with the results devoted to the three-dimensional
gas \cite{18}.

Let us start with statement problem.
Then we will give the solution of the  Smoluchowsky'
problem for the one-dimensional kinetic equation with frequency of collisions,
affine depending  on the module of molecular velocity.

\begin{center}
  {\bf 1. Statement of the problem and the basic equations}
\end{center}

Let us start with statement of a problem Smoluchowsky for
the one-dimensional kinetic equation with frequency of collisions,
affine  depending on the module velocity of molecules.

Let us begin with the general statement. Let gas occupies
half-space $x> 0$. The surface temperature $T_s $  and
concentration of sated steam of a surface $n_0 $ are set.
Far from a surface gas moves with some velocity $u $,
being velocity of evaporation (or condensation),
also has the temperature gradient
$$
g_T=\Big(\dfrac{d\ln T}{dx}\Big)_{x=+\infty}.
$$

It is necessary to define jumps of
temperature and concentration depending on velocity and
temperature gradient.

In a problem about weak evaporation
it is required to define tempe\-ra\-ture and concentration jumps
depending on velocity, including a tem\-pe\-ra\-ture gradient
equal to zero, and velocity of evaporation (condensation) is enough small.
The last means, that
$$
u \ll v_T.
$$

Here $v_T$ is the heat velocity of molecules, having order
of sound velocity order,
$$
v_T=\dfrac{1}{\sqrt{\beta_s}}, \qquad \beta_s=\dfrac{m}{2k_BT_s},
$$
$m$ is the mass of molecule, $k_B$ is the Boltzmann constant.

In the problem about tem\-pe\-rature jump it is required to define
tem\-pe\-rature and concentration jumps
depending on a temperature gradient, thus
evaporation (condensation) velocity
it is considered equal to zero, and the temperature gradient is
considered  as small.
It means, that
$$
lg_T\ll 1, \qquad l=\tau v_T,\qquad \tau=\dfrac{1}{\nu_0},
$$
where $l$ is the mean free path of gas molecules,
$\tau$ is the mean relaxation time, i.e. time between two
consecutive collisions of molecules.

Let us unite both problems  (about weak evaporation (condensation) and
temperature jump) in one. We will assume that the gradient
of temperature is small (i.e. relative difference
of temperature on length of mean free path is small) and the
velocity of gas in comparison with sound velocity is small. In
this case the problem supposes linearization and  distribution function
it is possible to search in the form
$$
f(x,v)=f_0(v)(1+h(x,v)),
$$
where
$$
f_0(v)=n_s\Big(\dfrac{m}{2\pi k_BT_s}\Big)^{1/2}
\exp \Big[-\dfrac{mv^2}{2k_BT_s}\Big]
$$
is the absolute Maxwellian.

Let us pass in the equation (1.1) to dimensionless velocity
$$
C=\sqrt{\beta}v=\dfrac{v}{v_T}
$$
and dimensionless coordinate
$$
x'=\nu_0 \sqrt{\dfrac{m}{2k_BT_s}}x=\dfrac{x}{l}
$$

The variable $x'$ let us designate again through $x$.

We take the linear kinetic equation  \cite{1}
$$
\mu\dfrac{\partial h}{\partial x}+(1+\sqrt{\pi}a|\mu|)h(x,\mu)=
$$
$$
=(1+\sqrt{\pi}a|\mu|)\dfrac{1}{\sqrt{\pi}}\int\limits_{-\infty}^{\infty}
e^{-\mu'^2}(1+\sqrt{\pi}a|\mu'|)q(\mu,\mu',a)h(x,\mu')d\mu'.
\eqno{(1.1)}
$$

Here $q(\mu,\mu',a)$ is the kernel of equation, \medskip
$$
q(\mu,\mu',a)=r_0(a)+r_1(a)\mu\mu'+r_2(a)(\mu^2-\beta(a))(\mu'^2-\beta(a)),
$$ \medskip
$$
r_0(a)=\dfrac{1}{a+1},\qquad r_1(a)=\dfrac{2}{2a+1},\qquad
r_2(a)=\dfrac{4(a+1)}{4a^2+7a+2},
$$
$$
\beta(a)=\dfrac{2a+1}{2(a+1)},
$$
$a$ is the  arbitrary positive paramater, $0\leqslant a<+\infty$.

Let us notice, that at $a\to 0$ the equation (1.1)
passes in the equation

$$
\mu\dfrac{\partial h}{\partial x}+h(x,\mu)=\dfrac{1}{\sqrt{\pi}}
\int\limits_{-\infty}^{\infty}e^{-\mu'^2}q(\mu,\mu')h(x,\mu)d\mu
\eqno{(1.2)}
$$
with kernel
$$
q(\mu,\mu')=1+2\mu\mu'+2\Big(\mu^2-\dfrac{1}{2}\Big)
\Big(\mu'^2-\dfrac{1}{2}\Big).
$$

This equation is one-dimensional BGK-equation with  constant
frequ\-ency of collisions.

Let us consider the second limiting case of the equation (1.1).
We will return to  expression of frequency of collisions
also we will copy it in the form
$$
\nu(\mu)=\nu_0(1+\sqrt{\pi}a|\mu|)=\nu_0+\nu_1|\mu|,
$$
where
$$
\nu_1=\sqrt{\pi}\nu_0 a.
$$

Let us tend $ \nu_0$ to zero. In this limit the quantity $a $ tends to
$ + \infty $, because
$$
a=\dfrac{\nu_1}{\sqrt{\pi}\nu_0}.
$$

It is easy to see, that in this limit
$$
\lim\limits_{a\to+\infty}(1+\sqrt{\pi}a|C'|)q(\mu,\mu',a)=
\sqrt{\pi}|\mu'|q_1(\mu,\mu'),
$$
where
$$
q_1(\mu,\mu')=1+\mu\mu'+(\mu^2-1)(\mu'^2-1).
$$

The equation (1.1) will thus be copied in the form
$$
\dfrac{\mu}{|\mu|}\dfrac{\partial h}{\partial x_1}+h(x_1,\mu)=
$$
$$=
\int\limits_{-\infty}^{\infty}e^{-\mu'^2}|\mu'|
[1+\mu\mu'+(\mu^2-1)(\mu'^2-1)]d\mu'.
\eqno{(1.3)}
$$
In this equation
$$
x_1=\nu_1\sqrt{\beta_s}x=\dfrac{x}{l_1},\qquad l_1=v_T\tau_1,\qquad
\tau_1=\dfrac{1}{\nu_1}.
$$

This equation is the one-dimensional kinetic equation with
the frequ\-ency of collisions proportional to the module of the molecular
velocity.

\begin{center}
  \bf 2. Kinetic equation with constant collision frequency.
Statement of boundary problem
\end{center}

Rectilinear substitution it is possible to check up, that the kinetic
equation (1.1) has following four private solutions
$$
h_0(x,\mu)=1,
$$
$$
h_1(x,\mu)=\mu,
$$
$$
h_2(x,\mu)=\mu^2,
$$
$$
h_3(x,\mu)=\Big(\mu^2-\dfrac{3}{2}\Big)\Big(x-\mu).
$$

Let us consider, that molecules are reflected from a wall purely
dif\-fu\-si\-vely, i.e. they are reflected from a wall with Maxwell distribution
by velocities, i.e.
$$
f(x,v)=f_0(v),\qquad v_x>0.
$$
From here we receive for function $h(x,C)$ condition
$$
h(0,\mu)=0, \qquad \mu>0.
\eqno{(2.1)}
$$

Condition (2.1) is the first boundary condition to the equation (1.2).

For asymptotic distribution of Chapmen---Enskog we will search in
the form of a linear combination of its partial solutions with
unknown coeffitients
$$
h_{as}(x,\mu)=A_0+A_1\mu+A_2\Big(\mu^2-\dfrac{1}{2}\Big)+
A_3\Big(\mu^2-\dfrac{3}{2}\Big)(x-\mu).
\eqno{(2.2)}
$$

We consider the distribution of number density
$$
n(x)=\int\limits_{-\infty}^{\infty}f(x,v)dv=
\int\limits_{-\infty}^{\infty}f_0(v)(1+h(x,v))dv=
n_0+\delta n(x).
$$

Here
$$
n_0=\int\limits_{-\infty}^{\infty}f_0(v)dv,\qquad
\delta n(x)=\int\limits_{-\infty}^{\infty}f_0(v)h(x,v)dv.
$$
From here we receive that
$$
\dfrac{\delta n(x)}{n_0}=\dfrac{1}{\sqrt{\pi}}
\int\limits_{-\infty}^{\infty}e^{-\mu^2}h(x,\mu)d\mu.
$$

We denote
$$
n_e=n_0\dfrac{1}{\sqrt{\pi}}
\int\limits_{-\infty}^{\infty}e^{-\mu^2}(1+h_{as}(x=0,\mu))d\mu.
$$

From here we receive that
$$
\varepsilon_n\equiv \dfrac{n_e-n_0}{n_0}=\dfrac{1}{\sqrt{\pi}}
\int\limits_{-\infty}^{\infty}e^{-\mu^2}h_{as}(x=0,\mu)d\mu.
\eqno{(2.3)}
$$

The quantity $\varepsilon_n$ is the unknown  jump of concentration.

Substituting (2.2) in (2.3), we find, that
$$
\varepsilon_n=A_0.
\eqno{(2.4)}
$$

From definition of dimensional velocity of gas
$$
u(x)=\dfrac{1}{n(x)}\int\limits_{-\infty}^{\infty}f(x,v)vdv
$$
we receive, that in linear approximation dimensional mass velocity
is equal
$$
U(x)=\dfrac{1}{\sqrt{\pi}}\int\limits_{-\infty}^{\infty}
e^{-\mu^2}h(x,\mu)\mu d\mu.
$$
Setting "far from a wall"\, velocity of evaporation (condensation),
let us write
$$
U=\dfrac{1}{\sqrt{\pi}}\int\limits_{-\infty}^{\infty}
e^{-\mu^2}h_{as}(x,\mu)\mu d\mu.
\eqno{(3.5)}
$$

Substituting in (2.5) distribution (2.2), we receive, that
$$
A_1=2U.
\eqno{(2.6)}
$$

Let us consider temperature distribution
$$
T(x)=\dfrac{2}{kn(x)}\int\limits_{-\infty}^{\infty}\dfrac{m}{2}
(v-u_0(x))^2f(x,v)dv.
$$

From here we find, that
$$
\dfrac{\delta T(x)}{T_0}=-\dfrac{\delta n}{n_0}+\dfrac{2}{\sqrt{\pi}}
\int\limits_{0}^{\infty}e^{-\mu^2}h(x,\mu)\mu^2d\mu=
$$
$$
=\dfrac{2}{\sqrt{\pi}}
\int\limits_{0}^{\infty}e^{-\mu^2}h(x,\mu)(\mu^2-\dfrac{1}{2})d\mu.
$$

From here follows, that at $x\to + \infty $ asymptotic
distribution is equal
$$
\dfrac{\delta T_{as}(x)}{T_0}=\dfrac{2}{\sqrt{\pi}}
\int\limits_{0}^{\infty}e^{-\mu^2}h_{as}(x,\mu)(\mu^2-\dfrac{1}{2})d\mu.
\eqno{(2.7)}
$$

Setting of the gradient of temperature far from a wall means, that
distribution of temperature looks like
$$
T(x)=T_e+\Big(\dfrac{dT}{dx}\Big)_{x=+\infty}\cdot x=T_e+G_Tx,
$$
where
$$
G_T=\Big(\dfrac{dT}{dx}\Big)_{+\infty}.
$$

This distribution we will present in the form
$$
T(x)=T_s\Big(\dfrac{T_e}{T_s}+g_Tx\Big)=T_s\Big(1+
\dfrac{T_e-T_s}{T_s}+g_Tx\Big), \quad x\to +\infty,
$$
where
$$
g_T=\Big(\dfrac{d\ln T}{dx}\Big)_{x=+\infty},
$$
or
$$
T(x)=T_s(1+\varepsilon_T+g_Tx),\qquad x\to +\infty,
$$
where
$$
\varepsilon_T=\dfrac{T_e-T_s}{T_s}
$$
is the unknown temperature jump.

From expression (2.7) is visible, that relative change
of temperature far from a wall is described by linear function
$$
\dfrac{\delta T_{as}(x)}{T_s}=\dfrac{T(x)-T_s}{T_s}=\varepsilon_T+g_Tx,\quad
x\to+\infty
\eqno{(2.8)}
$$

Substituting (2.2) in (2.7), we receive, that
$$
\dfrac{\delta T_{as}(x)}{T_s}=A_2+A_3x.
\eqno{(2.10)}
$$

Comparing (2.7) and (2.10), we find
$$
A_2=\varepsilon_T, \qquad A_3=g_T.
$$

So, asymptotic function of  Chapmen---Enskog' distribution
is const\-ruc\-ted
$$
h_{as}(x,\mu)=\varepsilon_n+ \varepsilon_T+2U\mu+
\Big(\mu^2-\dfrac{3}{2}\Big)[\varepsilon_T+g_T(x-\mu)].
$$

Now we will formulate the second boundary condition to the equation (1.2)
$$
h(x,\mu)=h_{as}(x,\mu)+o(1), \qquad x\to +\infty.
\eqno{(2.11)}
$$

Now we will formulate the basic boundary problem, which is gene\-ralized
Smoluchowsky' problem. This problem consists in finding of  such
solution of the kinetic equation (2.2) which satisfies
to boundary conditions (2.1) and (2.11).

\begin{center}
  \bf 3. Eigenvalues and eigenfunctions
\end{center}

Seperation of variables in the equation (1.2), taken in the form
Разделение переменных в уравнении (1.2), взятое в виде
$$
h_\eta(x,\mu)=\exp\Big(-\dfrac{x}{\eta}\Big)\Phi(\eta,\mu), \qquad
\eta \in \mathbb{C},
\eqno{(3.1)}
$$
reduces this equation to the characteristic
$$
(\eta-\mu)\Phi(\eta,\mu)=\dfrac{\eta}{\sqrt{\pi}}n_0(\eta)
+\dfrac{2\eta}{\sqrt{\pi}}\mu n_1(\eta)+
\dfrac{2\eta}{\sqrt{\pi}}\Big(\mu^2-\dfrac{1}{2}\Big)n_2(\eta),
\eqno{(3.2)}
$$
where
$
\eta,\mu\in (-\alpha,+\alpha),
$
$$
n_0(\eta)=\int\limits_{-\infty}^{\infty}e^{-\mu'^2}\Phi(\eta,\mu)d\mu,\qquad
n_1(\eta)=\int\limits_{-\infty}^{\infty}e^{-\mu'^2}\mu\Phi(\eta,\mu)d\mu,
$$
$$
n_2(\eta)=\int\limits_{-\infty}^{\infty}e^{-\mu'^2}
\mu^2\Phi(\eta,\mu)d\mu
$$
are the zeroes, first  and second moments of eigenfunction $\Phi(\eta,\mu)$.

Multiplying the characteristic equation (3.1) on $e^{-\mu'^2} $ and
integrating on all real axis, we receive, that
$$
n_1(\eta)\equiv 0.
$$
Multiplying the characteristic equation (3.1) on $\mu'e^{-\mu'^2} $ and
integrating on all real axis, we receive, that
$$
n_2(\eta)\equiv 0.
$$

We obtain the characteristic equation
$$
(\eta-\mu)\Phi(\eta,\mu)=\dfrac{\eta}{\sqrt{\pi}}\Big(\dfrac{3}{2}-\mu^2\Big)
\int\limits_{-\infty}^{\infty}e^{-\mu^2}\Phi(\eta,\mu)d\mu.
$$

Let us accept further the following normalization condition
for the eigenfunctions $\Phi(\eta,\mu)$:
$$
n_0(\eta) \equiv \int\limits_{-\infty}^{\infty}e^{-\mu'^2}\Phi(\eta,\mu)d\mu=1.
\eqno{(3.3)}
$$

Now the characteristic equation becomes
$$
(\eta-\mu)\Phi(\eta,\mu)=
\dfrac{\eta}{\sqrt{\pi}}\Big(\dfrac{3}{2}-\mu^2\Big).
\eqno{(3.4)}
$$

Eigenfunctions of the continuous spectrum filling
by  the continuous fashion the interval $ (-\infty, \infty) $,
We find \cite{19} in space of the generalized functions
$$
\Phi(\eta,\mu)=\dfrac{\eta}{\sqrt{\pi}}
\Big(\dfrac{3}{2}-\mu^2\Big)P\dfrac{1}{\eta-\mu}+
e^{\eta^2}\lambda(\eta)\delta(\eta-\mu),
\quad \eta\in (-\infty,\infty).
\eqno{(3.5)}
$$

Here $\lambda(\eta)$ is the dispersion fuction, defined by
equation (3.3), $Px^{-1}$  is the distribution, meaning
principal value of integral at intrgration $x^{-1}$,
$\delta(x)$ is the Dirac function,
$$
\lambda(z)=1+\dfrac{z}{\sqrt{\pi}}\int\limits_{-\infty}^{\infty}e^{-\tau^2}
\dfrac{3/2-\tau^2}{\tau-z}d\tau=-\dfrac{1}{2}+\Big(\dfrac{3}{2}-z^2\Big)
\lambda_0(z),
$$
$$
\lambda_0(z)=1+\dfrac{z}{\sqrt{\pi}}\int\limits_{-\infty}^{\infty}
e^{-\tau^2}\dfrac{d\tau}{\tau-z}=\dfrac{1}{\sqrt{\pi}}
\int\limits_{-\infty}^{\infty}
e^{-\tau^2}\tau\dfrac{d\tau}{\tau-z}.
$$

Apparently from the solution of the characteristic equation, continuous
spectrum of the characteristic equation is the set
$$
\sigma_\mu=\{\eta: -\infty<\eta<+\infty\}.
$$

By definition by the discrete spectrum of the characteristic equation
is set of zero of dispersion function.

Expanding dispersion function  in Laurent series in a vicinity
infinitely remote point, we are convinced, that it in this point
has zero of the fourth order. Applying an argument principle
from the theory of functions complex variable, it is possible to show, that
other zero, except $z_i =\infty $, dispersion function not
has. Thus, the discrete spectrum of the characteristic
equations consists of one point $z_i =\infty $, which multiplicity
is equal four,
$$
\sigma_d=\{z_i=\infty\}.
$$

To point $z_i =\infty $, as to a 4-fold point of the discrete spectrum,
corresponds the following four discrete (partial) solutions
the kinetic decision (1.2): \, $h_0 (x, \mu) $, \, $h_1 (x, \mu) $, \,
$h_3 (x, \mu) $ and $h_3 (x, \mu) $.

Let us result  Sokhotsky formulas for the difference and the sum of the boundary
values of dispersion function from above and from below on the cut
$(-\infty,+\infty)$:
$$
\lambda^+(\mu)-\lambda^-(\mu)=2\sqrt{\pi} i \mu e^{-\mu^2}
\Big(\dfrac{3}{2}-\mu^2\Big),\quad \mu\in (-\infty,+\infty),
$$
and
$$
\dfrac{\lambda^+(\mu)+\lambda^-(\mu)}{2}=
-\dfrac{1}{2}+\Big(\dfrac{3}{2}-\mu^2\Big)\lambda_0(\mu),\quad
\mu\in (-\infty,+\infty).
$$

On the real axis function $ \lambda_0 (\mu) $ is calculated on
to the formula
$$
\lambda_0(\mu)=1-2\mu^2\int\limits_{0}^{1}e^{-\mu^2(1-t^2)}dt.
$$

\begin{center}
  \bf 4. Homogeneous boundary Riemann problem
\end{center}

Here we will consider homogeneous boundary Riemann problem from
theories of functions complex variable which is required
further. This problem consists in the finding of such function $X(z)$,
which is analytical in a complex plane,
cut along the real positive half-axis
$ \mathbb{C}'=\mathbb{C}\setminus \mathbb {R}^+ $.

Boundary values of this function from above and from below on
the real half-axis satisfy to the boundary condition
$$
\dfrac{X^+(\mu)}{X^-(\mu)}=\dfrac{\lambda^+(\mu)}{\lambda^-(\mu)},\qquad
\mu>0.
\eqno{(4.1)}
$$

We note that
$$
|\lambda^+(\mu)|=|\lambda^-(\mu)|, \qquad
\lambda^+(\mu)=\overline{\lambda^-(\mu)},\qquad
\mu\in(-\infty,+\infty).
$$

Let us enter the  principal value of argument $\theta (\mu) = \arg
\lambda^+(\mu) $, defined in the cut plane
$ \mathbb {C}'$ and fixed in zero by the condition $ \theta(0) =0$.
Then
$$
\lambda^+(\mu)=|\lambda^+(\mu)|e^{i \theta(\mu)},\qquad
\lambda^-(\mu)=|\lambda^-(\mu)|e^{-i \theta(\mu)}.
$$

Noe the problem (4.1)  will be rewritten in the form
$$
\dfrac{X^+(\mu)}{X^-(\mu)}=e^{2i \theta(\mu)}, \qquad \mu>0.
\eqno{(4.2)}
$$

Taking the logarithm of the problem (4.2), we receive following nume\-rable
family of problems of the finding of analytical function on its zero
jump on the positive real half-axis $\mathbb{R}^+=
\{\mu: \mu>0\}$:
$$
\ln X^+(\mu)-\ln X^-(\mu)=2i \theta(\mu)+2\pi i k,\quad k\in \mathbb{Z},\quad
\mu>0.
\eqno{(4.3)}
$$

\begin{figure}[th]
\begin{center}
\includegraphics[width=16.0cm, height=10cm]{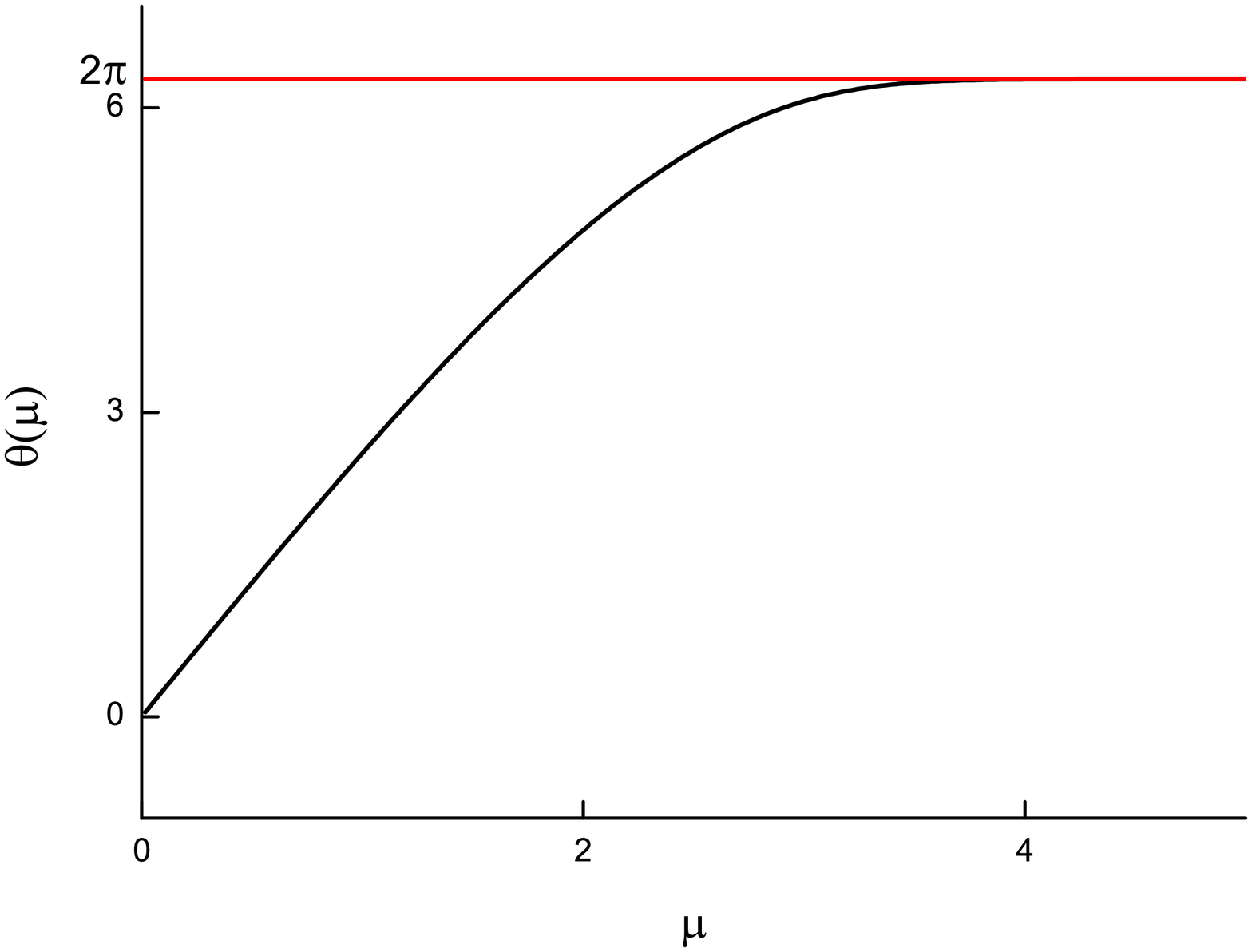}
\end{center}
\center{Рис. 1. The angle $\theta=\theta(\mu)$ monotonously increases
from $0$ to $2\pi$.}
\end{figure}

The solution of problems (4.3) is expressed by integral of Cauchy type
$$
\ln X(z)=\dfrac{1}{\pi}\int\limits_{0}^{\infty}
\dfrac{\theta(\mu)+k \pi}{\mu-z}d\mu, \qquad  k\in \mathbb{Z}.
$$

Let us notice, that the angle $ \theta(\mu)$ is on a semiaxis
$\mathbb{R}^+$ monotonously increasing function from $0$ to 2$ \pi $.
It means, that index  of coefficient $G(\mu)=\dfrac{\lambda^+(\mu)}
{\lambda^-(\mu)}$ of homogeneous Riemann problem
(5.1) on the positive real half-axis is equal to unit
$$
\varkappa=\varkappa(G)=\dfrac{1}{2\pi}
\Big[\arg G(\mu)\Big]\Big|_{0}^\infty=1.
$$

From here follows, that among family of solutions (4.3) only one (at $k =-2$)
is expressed by the converging integral of Cauchy type
$$
\ln X(z)=\dfrac{1}{\pi}\int\limits_{0}^{\infty}
\dfrac{\theta(\mu)-2 \pi}{\mu-z}d\mu.
\eqno{(4.4)}
$$

We denote further
$$
V(z)=\ln X(z),
$$
whence
$$
X(z)=e^{V(z)}.
$$

Let us redefine the received solution as follows
$$
X(z)=\dfrac{1}{z^2}e^{V(z)}.
\eqno{(4.5)}
$$

Let us notice, that the solution (4.5) isbounded function in a vicinity
the point $z=0$. Really, at $z\to 0$ it is had
$$
V (z) =-\dfrac {\theta (0)-2\pi} {\pi} \ln z+O (z), \qquad z\to 0,
$$
where $O(z)$ is the bounded function in a vicinity
the point $z=0$.
Hence, in a vicinity  of the point $z=0$  function
$X(z)=e^{O(z)}$ is  the bounded function.

\begin{center}
  \bf 5. Analytical splution of the boundary problem for
kinetic equation with constant collision frequency
\end{center}

Here we will prove the theorem about the analytical solution of
the basic boundary problem (1.2), (2.1) and (2.11).

{\sc Theorem.} {\it Boundary problem (1.2), (2.1) and (2.11) has
the unique decision, representable in the form of the sum linear
combinations of discrete (partial) solutions of this equation
and integral on the continuous spectrum
from eigenfunctions correponding to the continuous spectrum
$$
h(x,\mu)=h_{as}(x,\mu)+\int\limits_{0}^{\infty}
\exp\Big(-\dfrac{x}{\eta}\Big)\Phi(\eta,\mu)A(\eta)d\eta.
\eqno{(5.1)}
$$

In equality (5.1) $ \varepsilon_n $ and $ \varepsilon_T $ are
unknown coefficient (discrete spectrum), $U $ and $g_T $
are the given qualities, $A(\eta)$ is the unknown function
(coefficient of the continuous spectrum).}

Coefficients of discrete and continuous spectra are subject
to finding from boundary conditions.

Expansion (5.1) it is possible to present in the explicit form in
classical sense
$$
h(x,\mu)=\varepsilon_n+\varepsilon_T+2U\mu+
\Big(\mu^2-\dfrac{3}{2}\Big)[\varepsilon_T+g_T(x-\mu)]+
$$
$$
+e^{\mu^2-x/\mu}\lambda(\mu)A(\mu)+\Big(\dfrac{3}{2}-\mu^2\Big)
\dfrac{1}{\sqrt{\pi}}\int\limits_{0}^{\infty}
e^{-x/\eta}\dfrac{\eta A(\eta)}{\eta-\mu}d\eta.
\eqno{(5.1')}
$$

{\sc Proof.}
Let us substitute expansion (5.1) in the boundary condition (2.1).
We receive the integral equation
$$
h_{as}(0,\mu)+\int\limits_{0}^{\infty}\Phi(\eta,\mu)A(\eta)d\eta=0,\quad
0<\mu<\infty.
$$

In the explicit form this equation looks like
$$
h_{as}(0,\mu)+e^{\mu^2}\lambda(\mu)A(\mu)+
$$
$$
+\Big(\dfrac{3}{2}-\mu^2\Big)
\dfrac{1}{\sqrt{\pi}}\int\limits_{0}^{\infty}
\dfrac{\eta A(\eta)}{\eta-\mu}d\eta=0,\qquad 0<\mu<\infty.
\eqno{(5.2)}
$$

Here
$$
h_{as}(0,\mu)=\varepsilon_n+\varepsilon_T+2U\mu+
\Big(\mu^2-\dfrac{3}{2}\Big)(\varepsilon_T-g_T\mu).
$$

Let us enter auxiliary function
$$
N(z)=\dfrac{1}{\sqrt{\pi}}\int\limits_{0}^{\infty}
\dfrac{\eta A(\eta)}{\eta-z}d\eta,
\eqno{(5.3)}
$$
for which according to formulas Sokhotsky it is had
$$
N^+(\mu)-N^-(\mu)=2\sqrt{\pi} i \mu A(\mu),\quad 0<\mu<\infty,
\eqno{(5.4)}
$$
$$
\dfrac{N^+(\mu)+N^-(\mu)}{2}=\dfrac{1}{\sqrt{\pi}}\int\limits_{0}^{\infty}
\dfrac{\eta A(\eta)}{\eta-\mu}d\eta, \quad 0<\mu<\infty.
\eqno{(5.5)}
$$

Let us transform the equation (5.2), considering formulas Sokhotsky for
dispersion function and according to equalities (5.4) and (5.5).
We receive non-uniform the boundary Riemann condition

Considering formulas Sokhotsky for dispersive function,
Let's transform the equation (5.6) to a non-uniform regional problem
Римана:
$$
\lambda^+(\mu)\Big[\Big(\dfrac{3}{2}-\mu^2\Big)N^+(\mu)+h_{as}(0,\mu)\Big]-
$$
$$
-\lambda^-(\mu)\Big[\Big(\dfrac{3}{2}-\mu^2\Big)N^-(\mu)+h_{as}(0,\mu)\Big]=0,
\quad 0<\mu<\infty.
\eqno{(5.6)}
$$

Let us consider the corresponding homogeneous boundary Riemann problem
$$
\dfrac{X^+(\mu)}{X^-(\mu)}=\dfrac{\lambda^+(\mu)}{\lambda^-(\mu)},
\quad 0<\mu<\infty.
\eqno{(5.7)}
$$

The solution of this problem which is bounded and not disappearing
in points $z=0$ and $z =\alpha $ it is resulted in the previous item
$$
X(z)=\dfrac{1}{z^2}\exp V(z),
\eqno{(5.8)}
$$
where
$$
V(z)=\dfrac{1}{\pi}\int\limits_{0}^{\infty}\dfrac{\theta(\mu)-2\pi}
{\mu-z}d\mu,
\eqno{(6.9)}
$$
$\theta(\mu)=\arg \lambda^+(\mu)$ is the principal value of argument,
fixed by condition $\theta(0)=0$.

Let us transform the problem (5.6) by means of the homogeneous problem (5.7) to
the problem of finding of analytical function on its jump on the cut
$$
X^+(\mu)\Big[\Big(\dfrac{3}{2}-\mu^2\Big)N^+(\mu)+h_{as}(0,\mu)\Big]=
$$
$$
=X^-(\mu)\Big[\Big(\dfrac{3}{2}-\mu^2\Big)N^-(\mu)+h_{as}(0,\mu)\Big],
\quad 0<\mu<\infty.
\eqno{(5.10)}
$$

Let us find singularities of the boundary condition (5.10).
Considering behaviour of the functions entering into boundary
condition (5.10), we receive the common solution corresponding
to boundary problem
$$
\Big(z^2-\dfrac{3}{2}\Big)N(z)=h_{as}(0,z)+\dfrac{C_0+C_1z}{X(z)},
\eqno{(5.11)}
$$
where $C_0$ and $C_1$ are arbitrary constans, and
$$
h_{as}(0,z)=\varepsilon_n+\varepsilon_T+2U\mu+
\Big(z^2-\dfrac{3}{2}\Big)(\varepsilon_T-g_Tz).
$$

Let us notice, that the solution (5.11) has in infinitely removed
point $z =\infty $ a pole of the third order, while function
$N(z)$, defined by equality (5.3), has in this point a pole
the first order.

That the solution (5.11) could be accepted in
quality of function $N(z)$, defined by equality (5.3), we will lower
order of a pole at the solution (5.11) from three to unit.

Then let us equate values of the left and right parts of equality
(5.11) in points of the real axis $\mu_{1,2}=\pm \sqrt{3/2} $.

Decomposition is required to us
$$
V(z)=\dfrac{V_1}{z}+\dfrac{V_2}{z^2}+\cdots, \qquad z\to \infty.
$$

Here
$$
V_n=-\dfrac{1}{\pi}\int\limits_{0}^{\infty}\tau^{n-1}[\theta(\tau)-
2\pi]d\tau, \qquad n=1,2,\cdots.
$$

Lowering an order of pole on two units in infinitely removed
point at the soltution (5.11), we find, that
$$
C_0= V_1g_T-\varepsilon_T,
$$
$$
C_1=g_T.
$$

The pole of function in the point $ \mu_1 =\sqrt{3/2} $ is eliminated
by two limiting conditions from above and from below the real axis,
for this point lays on a cut (the real axis)
$$
C_0+C_1\mu_1+X^+(\mu_1)(\varepsilon_n+\varepsilon_T+2U\mu_1)=0,
\eqno{(5.12)}
$$
and
$$
C_0+C_1\mu_1+X^-(\mu_1)(\varepsilon_n+\varepsilon_T+2U\mu_1)=0,
\eqno{(5.13)}
$$

The point $\mu_2=-\mu_1$ does not belong to the cut, therefore we
receive
$$
C_0-C_1\mu_1+X(-\mu_1)(\varepsilon_n+\varepsilon_T-2U\mu_1)=0.
\eqno{(5.14)}
$$

We take a half-sum of conditions (5.12) and (5.13)
$$
C_0+C_1\mu_1+\dfrac{X^+(\mu_1)+X^-(\mu)}{2}
(\varepsilon_n+\varepsilon_T+2U\mu_1)=0
\eqno{(5.15)}
$$

We note that
$$
X^{\pm}(\mu_1)=\dfrac{1}{\mu_1^2}e^{V^{\pm}(\mu_1)},
$$
where
$$
V^{\pm}(\mu_1)=V(\mu_1)\pm i[\theta(\mu_1)-2\pi].
$$
Hence
$$
\dfrac{X^+(\mu_1)+X^-(\mu)}{2}=X(\mu_1).
$$

Taking into account this equality from the equations (5.14)
and (5.15) it is received expressions of required qualities of
jump of temperature and jump concentration
$$
\varepsilon_T=g_T\Big[V_1-\mu_1\dfrac{X(\mu_1)+X(-\mu_1)}
{X(\mu_1)-X(-\mu_1)}\Big]-4U\mu_1\dfrac{X(\mu_1)X(-\mu_1)}
{X(\mu_1)-X(-\mu_1)}
\eqno{(5.16)}
$$
and
$$
\varepsilon_n=g_T\Big[V_1+\mu_1\dfrac{X(\mu_1)+X(-\mu_1)-2}
{X(\mu_1)-X(-\mu_1)}\Big]- \hspace{3cm}
$$
$$
-2U\mu_1\dfrac{X(\mu_1)+X(-\mu_1)-2X(\mu_1)X(-\mu_1)}
{X(\mu_1)-X(-\mu_1)}.
\eqno{(5.17)}
$$

Coefficient of continuous spectrum $A(\eta $ can be found  on the basis
of the formula Sokhotsky (5.4) and
formulas of the difference of boundary values $N(z)$, received with the help
solutions (5.11)
$$
N+(\mu)-N^-(\mu)=\dfrac{C_0+C_1\mu}{\mu^2-{3}/{2}}\Big[
\dfrac{1}{X^+(\mu)}-\dfrac{1}{X^-(\mu)}\Big].
\eqno{(5.18)}
$$

From equalities (5.4) and (5.18) we find coefficient of the continuous
spectrum
$$
2\sqrt{\pi}i\eta A(\eta)=\dfrac{g_T(V_1+\eta)-\varepsilon_T}{\eta^2-3/2}
\Big[\dfrac{1}{X^+(\eta)}-\dfrac{1}{X^-(\eta)}\Big].
\eqno{(5.19)}
$$\medskip

We note that
$$
\dfrac{1}{X^+(\eta)}-\dfrac{1}{X^-(\eta)}=
-\dfrac{2i}{X(\eta)}\sin \theta(\eta).
$$

By means of this equality coefficient of the continuous spectrum (5.19)
it is definitively equal
$$
\eta A(\eta)=-\dfrac{(V_1+\eta)g_T-\varepsilon_T}
{\sqrt{\pi}X(\eta)(\eta^2-3/2)}\sin \theta(\eta).
$$

So, all coefficients of expansion (5.1) are established. On
to construction, expansion (5.1) satisfies to boundary conditions
(2.1) and (2.11). That fact, that expansion (5.1) satisfies
to the equation (1.2), it is checked directly.

Uniqueness of decomposition (5.1) is proved by a method from
the opposite. The theorem is proved.
\medskip

\begin{center}
  {\bf 6. Temperature jump and weak evaporation (condensation).
Numerical calculations}
\end{center}

Numerical calculations of coefficients $V_n $ lead to the following
results
$$
V_1=2.6470\cdots,\qquad V_2=2.5, \qquad V_3=3.7153\cdots,
$$
and also
$$
\mu_1=1.2247\cdots,\qquad X(\mu_1)=3.8483\cdots,\qquad X(-\mu_1)=0.1732\cdots.
$$

Now it is required to us following

{\sc Theorem.} {\it For dispersion function $ \lambda(z)$
takes place the following  factorization formula
$$
\lambda(z)=- \dfrac{3}{4}X(z)X(-z),\qquad z\in \mathbb{C'},
$$
$$
\lambda^+(\mu)=- \dfrac{3}{4}X^+(\mu)X(-\mu), \qquad \mu>0,
$$
$$
\lambda^-(\mu)=- \dfrac{3}{4}X(\mu)X^+(-\mu), \qquad \mu<0.
$$}

{\sc Proof.}  This theorem is proved in the same way, as well as
the proof of similar theorems in our works \cite{3}.

By means of this theorem it is found exact value
$$
X(\mu_1)X(-\mu_1)=\dfrac{2}{3}.
$$

We rewrite thiese formulas (5.16) and (5.17) in the form
$$
\varepsilon_T=K_{TT}g_T+K_{TU}(2U),\qquad
\varepsilon_n=K_{nT}g_T+K_{nU}(2U).
$$

Here
$$
K_{TT}=V_1-\mu_1\dfrac{X(\mu_1)+X(-\mu_1)}
{X(\mu_1)-X(-\mu_1)},
$$
$$
K_{TU}=-2\mu_1\dfrac{X(\mu_1)X(-\mu_1)}{X(\mu_1)-X(-\mu_1)},
$$
$$
K_{nT}=V_1+\mu_1\dfrac{X(\mu_1)+X(-\mu_1)-2}
{X(\mu_1)-X(-\mu_1)},
$$
$$
K_{nU}=-\mu_1\dfrac{X(\mu_1)+X(-\mu_1)-2X(\mu_1)X(-\mu_1)}
{X(\mu_1)-X(-\mu_1)}.
$$

Now it is easy to find that
$$
K_{TT}=1.3068,\qquad K_{TU}=-0.4443,
$$
$$
K_{nT}=3.3207, \qquad K_{nU}=-0.8958.
$$

Hence, coefficient of jump of temperature and jump
of concentration are calculated under formulas
$$
\varepsilon_T=1.3068g_T-0.4443(2U),
\eqno{(6.1)}
$$
and
$$
\varepsilon_n=-3.3207g_T-0.8958(2U).
\eqno{(6.2)}
$$

\begin{center}
  {\bf 7. Limiting transition in the general formulas}
\end{center}

Here we will show, that if we will make limiting transition in
the general formulas (7.8) and (7.9) from \cite{2} at $a\to 0$,
we in accuracy
let us receive formulas (5.16) and (5.17).
We will remind, that formulas (7.8) and
(7.9) for temperature and concentration jump are received for the case
frequencies of collisions, affine depending on the module of velocity.

We will do this transition for temperature jump in the case
of weak evaporations (i.e. for the case $g_T=0$).

Let us transform the formula (5.16) to the form
$$
\varepsilon_T=-2U\dfrac{1}{\dfrac{1}{2\mu_1X(-\mu_1)}-
\dfrac{1}{2\mu_1X(\mu_1)}}.
\eqno{(7.1)}
$$

From formula (7.8) from \cite{2} we receive
$$
\varepsilon_T=-2U\dfrac{1}{V_1+\lim\limits_{a\to 0}K_1},
\eqno{(7.2)}
$$
where
$$
K_1=\dfrac{1}{2\pi i}\int\limits_{0}^{1/a}\Big(\dfrac{1}{X^+(\mu)}-
\dfrac{1}{X^-(\mu)}\Big)\dfrac{d\mu}{Q(\mu,\mu)}.
$$

For coincidence of equalities (7.1) and (7.2) it is required to prove
equality
$$
\lim\limits_{a\to 0}K_1=-V_1+\dfrac{1}{2\mu_1X(-\mu_1)}-
\dfrac{1}{2\mu_1X(\mu_1)}
\eqno{(7.3)}
$$

We note that in considering case at $a\to 0$: $r_2(a)\to
2$, $\beta(a)\to \dfrac{1}{2}$, $\omega(a)\to 0$, $r_0(a)\to 1$,
$r_1(a)\to 2$. Hence,
$$
\Lambda_2(\mu) \equiv 0,\quad \Lambda_0(\mu)\equiv 1,\quad
Q(\mu,\mu)=\dfrac{3}{2}-\mu^2.
$$

Therefore
$$
\lim\limits_{a\to 0}=\dfrac{1}{2\pi i}\int\limits_{0}^{\infty}
\Big(\dfrac{1}{X^+(\mu)}-
\dfrac{1}{X^-(\mu)}\Big)\dfrac{d\mu}{Q(\mu,\mu)}.
$$

We form the difficult contour $ \Gamma_R $, consisting of the external
circles $ |z |=R $ with radius $R=1/a $, and two internal circles
$ |z\pm \mu_1 | = a $ with radiuses $r=a $.

Under Cauchy theorem
$$
\oint\limits_{\Gamma_R}\dfrac{d\mu}{X(\mu)Q(\mu,\mu)}=0.
$$

We denote
$$
\lim\limits_{a\to 0}K_1=I.
$$
Then from the previous equality it is received
$$
I=I_1+I_2-I_3,
$$
where
$$
I_1=\lim\limits_{a\to 0}\dfrac{1}{2\pi i}\oint\limits_{|\mu+\mu_1|=a}
\dfrac{d\mu}{X(\mu)Q(\mu,\mu)},
$$
$$
I_2=\lim\limits_{a\to 0}\dfrac{1}{2\pi i}\oint\limits_{|\mu-\mu_1|=a}
\dfrac{d\mu}{X(\mu)Q(\mu,\mu)},
$$
$$
I_3=\lim\limits_{a\to 0}\dfrac{1}{2\pi i}\oint\limits_{|\mu|=1/a}
\dfrac{d\mu}{X(\mu)Q(\mu,\mu)}.
$$

It is easy to see, that
$$
I_1=\dfrac{1}{Q'(\mu,\mu)X(\mu}\Bigg|_{\mu=-\mu_1}=\dfrac{1}{2\mu_1X(\mu_1)},
$$
in the same way
$$
I_2=-\dfrac{1}{2\mu_1X(\mu_1)}.
$$

For  calculation of integral $I_3$ we will spread out its subintegral
function by Laurent series in a vicinity of infinitely remote point and
let us present it in the form
$$
\dfrac{1}{Q(z,z)X(z)}=-1+\dfrac{V_1}{z}+\varphi(z),
\qquad \mu\to \infty,
$$
where
$$
\varphi(z)=O(z^{-2}), \qquad z\to \infty.
$$

Hence, this integral equals
$$
\dfrac{1}{2\pi i}\oint\limits_{|\mu|=1/a}
\dfrac{d\mu}{X(\mu)Q(\mu,\mu)}=V_1+\dfrac{1}{2\pi i}
\oint\limits_{|\mu|=1/a}\varphi(\mu)d\mu.
$$

The limit of last integral is equal in this equality to zero at $a\to
0$ owing to previous asmptotic, therefore $I_3=V_1$.

So, equality (7.3) is established.

\begin {center}
  \bf 8. Distribution of macroparameters of gas in "half-space"
\end {center}

Let us consider distribution of concentration, mass velocity and
tem\-pe\-ra\-ture depending on coordinate $x $.

Let us begin with concentration distribution (numerical density)
$$
\dfrac{\delta n(x)}{n_0}=\dfrac{1}{\sqrt{\pi}}\int\limits_{-\infty}^{\infty}
e^{-\mu^2}h(x,\mu)d\mu=
$$
$$
=\dfrac{1}{\sqrt{\pi}}\int\limits_{-\infty}^{\infty}e^{-\mu^2}
\Big[h_{as}(x,\mu)+\int\limits_{0}^{\infty}e^{-x/\eta}\Phi(\eta,\mu)
A(\eta)d\eta\Big]d\mu=
$$
$$
=\varepsilon_T-g_Tx+\dfrac{1}{\sqrt{\pi}}\int\limits_{0}^{\infty}
e^{-x/\eta}d\eta\int\limits_{-\infty}^{\infty}e^{-\mu^2}\Phi(\eta,\mu)A(\eta)
d\eta.
$$

Having taken advantage of the normalizing equality (3.3), we receive
$$
\dfrac{\delta n(x)}{n_0}=
\varepsilon_T-g_Tx+\dfrac{1}{\sqrt{\pi}}\int\limits_{0}^{\infty}
e^{-x/\eta}A(\eta)d\eta.
$$

Let us transform coefficient  of the continuous spectrum. Noticing, that
$$
\sin \theta(\eta) = \dfrac{\sqrt{\pi} \eta e^{-\eta^2}(3/2-\eta^2)}
{|\lambda^+(\eta)|}.
$$
Hence,
$$
A(\eta)=\dfrac {(V_1 +\eta) g_T-\varepsilon_T}
{X(\eta)|\lambda^+(\eta)|} e^{-\eta^2}.
$$

Thus, we come to following distribution
of concentration
$$
\dfrac{\delta n(x)}{n_0} = [K_{TT}(1-m_0 (x))-x+V_1m_0 (x)+m_1 (x)] g_T+
$$
$$
+K_{TU} (1-m_0(x))(2U).
$$

Here
$$
m_k(x)=\dfrac{1}{\sqrt{\pi}}\int\limits_{0}^{\infty}
\dfrac{e^{-x/\eta-\eta^2}\eta^kd\eta}{X(\eta)|\lambda^+(\eta)|},\qquad k=0,1.
$$

Mass velocity $U (x) $ is equal everywhere at $x> 0$ to given on
infinity quantity of velocity, i.e. $U (x) \equiv U $. Really,
we have
$$
U(x)=\dfrac{1}{\sqrt{\pi}}\int\limits_{-\infty}^{\infty}
e^{-\mu^2}h(x,\mu)\mu d\mu=
$$
$$
=\dfrac{1}{\sqrt{\pi}}\int\limits_{-\infty}^{\infty}
e^{-\mu^2}\Bigg[h_{as}(x,\mu)+\int\limits_{0}^{\infty}e^{-x/\eta}
\Phi(\eta,\mu)A(\eta)d\eta\Bigg]\mu d\mu.
$$

From here we obtain that
$$
U(x)=U+\int\limits_{0}^{\infty}e^{-x/\eta}A(\eta)d\eta
\int\limits_{-\infty}^{\infty}e^{-\mu^2}\mu \Phi(\eta,\mu)d\mu \equiv U,
$$
because the first moment of eigenfunction $\Phi(\eta,\mu)$ is equal
to zero as it has been shown above.

We consider the distribution of temperature
$$
\dfrac{\delta T(x)}{T_0}=\varepsilon_T+g_Tx+
\dfrac{2}{\sqrt{\pi}}\int\limits_{-\infty}^{\infty}e^{-\mu^2}\Big(
\mu^2-\dfrac{1}{2}\Big)h(x,\mu)d\mu=
$$
$$
=\varepsilon_T+g_Tx+
\dfrac{2}{\sqrt{\pi}}\int\limits_{0}^{\infty}e^{-x/\eta}A(\eta)d\eta
\int\limits_{-\infty}^{\infty}e^{-\mu^2}\Big(
\mu^2-\dfrac{1}{2}\Big)\Phi(\eta,\mu)d\mu.
$$

Considering, that the second moment of eigenfunction $\Phi(\eta,\mu) $
is equal to zero, from here we receive the temperature distribution
$$
\dfrac{\delta T(x)}{T_0}=\varepsilon_T+g_Tx-\dfrac{1}{\sqrt{\pi}}
\int\limits_{0}^{\infty}e^{-x/\eta}A(\eta)d\eta=
$$
$$
=[x-m_1(x)+K_{TT}(1+m_0(x))]g_T+K_{TU}(1+m_0(x))(2U).
$$

\begin{center}
  \bf 9. Conclusion
\end{center}

In the present work the analytical solution of
boundary problems for the one-dimensional kinetic equation with
constant frequency of collisions of molecules  is considered.
We consider the solution of the generalized Smoluchowsky
 problem (problems about temperature jump and weak
evaporation (condensation)). Numerical calculations are done.
Distribution of concentration, mass speed and temperature is
received.


\begin{thebibliography}{99}

\bibitem{1}
{\it Latyshev A.V., Yushkanov A.A.}
The kinetic one-dimensional equation with frequency of collisions,
affine depending on the module molecular velocity//
arXiv:1403.2068v1 [math-ph] 9 Mar 2014, 20pp.

\bibitem{2}
{\it Latyshev A.V., Yushkanov A.A.}
Boundary problems for the one-dimensional kinetic equation
with fre\-qu\-ency of collisions, affine depending on the module
velocity// ArXiv:1403. 5854, [math-ph] 23 Mar 2014, 30 pp.


\bibitem{3}
{\it Latyshev A.V., Yushkanov A.A.}
Kinetic equatios type Williams and their exact solutions.
Monograph. M.: MGOU (Moscow State Regional University), 2004, 271 p.

 \bibitem{4}{\it  Latyshev A.V.}
 Application of case' method to the solution of linear
kinetic BGK equation in a problem about temperature jump//
Appl. math. and mechanics. 1990. V. 54. Issue 4. P. 581--586.
[russian]

\bibitem{5}
{\it  Latyshev A.V., Yushkanov A.A. }
Boundary problems for model Boltzmann equation with
frequency proportional to velocity of moleculs//
Izvestiya Russian Academy of Science. Ser. Mechanika, Fluid
and Gas (Russian "Fluids Dynamics"). 1993.
\No 6. 143-155 pp. [russian]

\bibitem{6}
{\it  Latyshev A.V., Yushkanov A.A. }
Analytical solution of the problem
about strong evaporation (condensation)// Izvestiya Russian Academy of
Science. Ser. Mechanika, Fluid and Gas (Russian "Fluids Dynamics"). 1993.
\No 6. 143-155 pp. [russian]

\bibitem{7}
{\it cassell J.S., Williams M.M.R.}
An exact solution of the temperature slip problem in rarefied
gases// Transport Theory and Statist. Physics, 2(1), 81--90
(1972).

\bibitem{8} {\it Latyshev A.V., Yushkanov A.A. }
Temperature jump and weak evaporatuion in molecular gases//
J. of experim. and theor. physics. 1998.
 V. 114. Issue. 3(9). P. 956--971. [russian]

\bibitem{9} {\it Latyshev A.V., Yushkanov A.A. }
The Smoluchowski problem in polyatomic gases//
Letters in J. of Tech. Phys. 1998. V. 24. \No 17. P. 85--90. [russian]

\bibitem{10}{\it Latyshev A.V., Yushkanov A.A.}
Analytic solutions of boundary
value problem for model kinetic equatins// Math.
Models of Non-Linear Excitations, Transfer, Dynamics, and
control in condensed Systems and Other Media. Edited by
L.A. Uvarova and A.V. Latyshev. Kluwer Academic. New York--Moscow.
2001. P. 17--24.

\bibitem{11}
{\it Latyshev A.V., Yushkanov A.A.}
Smolukhowski problem for degenerate Bose
gases// Theoretical and Mathematical Physics. Springer New
York. Vol. 155, \No 3,June, 2008, pp. 936 -- 948.

\bibitem{12}
{\it Latyshev A.V., Yushkanov A.A.}
Temperature jump in degenerate quantum gases
in the presence of the Bose–Einstein condensate
// Theor. and Mathem. Phys. 2010. V. 162(1). P. 95--105 [russian]

\bibitem{13}
{\it Latyshev A.V., Yushkanov A.A.}
Temperature jump in degenerate quantum gases with the Bogoliubov
excitation energy and in the presence of the Bose–Einstein condensate,
Theoret. and Math. Phys., 165:1 (2010), 1358–1370.


\bibitem{14}
{\it Latyshev A.V., Yushkanov A.A.}
Smoluchowski problem for electrons in metal// Theor. and Mathem.
Phys. 2005, январь, Т. 142. \No 1. С. 93--111 [russian]

\bibitem{15}
{\it Latyshev A.V., Yushkanov A.A.}
Smoluchowski problem for metals with mirror-diffusive boundary conditions
//Theoretical and Mathematical Physics
October 2009, Volume 161, Issue 1, pp. 1403-1414.


\bibitem{16}
{\it Latyshev A.V., Yushkanov A.A.}
Boundary value problems for quantum gases. Monograph
M.: MGOU, 2012, 266 p.[russian]

\bibitem{17}
{\it Сercignani С., Frezzoti A.} Linearized analysis of a one-speed
B.G.K. model in the case of strong condensation// Bulgarian Academy
of sci. theor. appl. mech. Sofia. 1988. V.XIX. \No 3. 19-23 P.

\bibitem{18}
{\it Latyshev A.V., Yushkanov A.A.}
Analytical solution of one-dimensional problem about
moderate strong evaporation (and condensation) in half-space//
Appl. mech. and tech. physics. 1993. \No 1. 102-109 p. [russian]

\bibitem{19}
{\it Vladimirov V.S., Zharinov V.V.} Equations of mathematical physics.
M.: Fizmatlit. 2000. 399 с.[russian]

\bibitem{20}
{\it Gakhov F.D.} Boundary value problems. M.: Nauka. 640 p.[russian]

\end{thebibliography}
\end{document}